\documentclass[pdflatex,sn-nature]{sn-jnl}

\usepackage{graphicx}%
\usepackage{multirow}%
\usepackage{amsmath,amssymb,amsfonts}%
\usepackage{amsthm}%
\usepackage{mathrsfs}%
\usepackage[title]{appendix}%
\usepackage{xcolor}%
\usepackage{textcomp}%
\usepackage{manyfoot}%
\usepackage{booktabs}%
\usepackage{algorithm}%
\usepackage{algorithmicx}%
\usepackage{algpseudocode}%
\usepackage{listings}%

\theoremstyle{thmstyleone}

\theoremstyle{thmstyletwo}%

\theoremstyle{thmstylethree}%

\begin{document}

\title[Anthropomorphism in AI Companion Communities]{Anthropomorphism in AI Companion Communities: Age, Gender, and Emotional Correlates}

\author[1]{\fnm{Afia} \sur{Mubashir}}\email{mubashir.afia@gmail.com}
\equalcont{These authors contributed equally to this work.}

\author[1]{\fnm{Boden} \sur{Moraski}}\email{bodenmoraski@gmail.com}
\equalcont{These authors contributed equally to this work.}

\author[1]{\fnm{Stephanie} \sur{Choi}}\email{stephaniechoii17@gmail.com}
\equalcont{These authors contributed equally to this work.}

\author[3,4]{\fnm{Rose E.} \sur{Guingrich}}\email{r.e.guingrich@gmail.com}

\affil[1]{\orgname{Independent}}

\affil[2]{\orgname{Princeton University}}

\affil[3]{\orgname{Ethicom}}

\abstract{Artificial intelligence (AI) systems are increasingly integrated into daily life, with millions now using AI chatbots built on Large Language Models (LLMs) for companionship. Both humanlike AI qualities and user predispositions to anthropomorphize relate to social consequences, such as increased trust, social health benefits, and psychological harms. Populations such as children, older adults, or those with mental health vulnerabilities may be particularly susceptible to anthropomorphism and its detriments, but mixed findings complicate the role of demographics. We used publicly available Reddit data from three popular AI companion subreddits to assess relationships between gender, age, anthropomorphism, and elicited emotions, to better understand how different people perceive and are affected by AI companions. We investigated three questions: How do age and gender relate to anthropomorphization of AI?, How does emotional expression relate to anthropomorphization?, and How do age and gender moderate emotion–anthropomorphization relationships? We found that adults and women anthropomorphize AI chatbots more than teens and men, and that positive emotional expression, particularly joy, is positively associated with anthropomorphization, while neutrality is negatively associated with anthropomorphism. Both relationships were stronger in adults than teens. Our findings suggest that the tendency to anthropomorphize may be more broadly distributed across age groups than previously expected, thereby prompting the reevaluation of existing digital safety norms.}


\maketitle

\section{Introduction}\label{sec1}

Social companion chatbots such as Character.AI and Replika are intended to fulfill specific functions of social companionship, simulating emotional connection, friendship, or even romance through AI-generated conversation. Such chatbots are highly anthropomorphic, as research in anthropomorphic qualities in AI chatbots is purported to increase trust and perceived accuracy of information, which can benefit AI companies by increasing user adoption \citep{cohnBelievingAnthropomorphismExamining2024}. Furthermore, recent research demonstrates chatbot users reported social companion chatbot use as beneficial to their social health and that an increased level of anthropomorphism was linked with stronger social health benefits and a more positive outlook on the chatbots \cite{guingrichChatbotsSocialCompanions2025}. Such anthropomorphic qualities possess several detriments, however: anthropomorphic chatbots can foster unhealthy emotional dependency, lead to social isolation, and promote harmful content \cite{namvarpourAIinducedSexualHarassment2025}. Populations such as children, older adults, or those suffering from mental illness may be particularly susceptible to these adverse effects. Notable media cases include 14-year-old Sewell Setzer III, who committed suicide after conversing with a chatbot \cite{rooseCanAIBe2024}; 19-year-old Chail, who attempted to assassinate Queen Elizabeth after conversations with an AI chatbot \cite{singletonHowChatbotEncouraged2023}; and 56-year-old Stein-Erik Solberg, who murdered his mother in a murder-suicide on her estate after talking to ChatGPT \cite{ruwitchNewLawsuitBlames2025}.

\subsection{Anthropomorphism} Anthropomorphism refers to the attribution of mental states or humanlike attributes to a non-human entity \cite{epleySeeingHumanThreefactor2007}. However, the traditional interpretation of anthropomorphism largely conceptualises it as a one-sided projection fallacy \cite{peterBenefitsDangersAnthropomorphic2025}. This framework becomes insufficient in the face of the fact that the latest generation of LLMs possess communicative abilities that approximate those of humans and behave “anthropomorphically” \cite{peterBenefitsDangersAnthropomorphic2025}. Previous work has focused on how anthropomorphism is moderated through cognitive and motivational determinants \cite{epleySeeingHumanThreefactor2007}, defining how humans understand what it's like to be an agent, predict its behavior, and connect with it. These factors build towards the idea that anthropomorphism increases when individuals seek relational fulfillment, predictive control or rely on accessible human schemas. Users with stronger social needs or relational interdependence may ascribe greater humanlike mind traits to AI systems, whereas individuals with higher self-efficacy or lower relational motivation may interpret the same system in more mechanistic terms \cite{guingrichLongitudinalRandomizedControl2025,epleyCreatingSocialConnection2008}, therefore reinforcing the fact that anthropomorphism is highly identity driven and anthropomorphic tendency may appear in varying degrees across demographics due to variance in social needs.

\subsection{Demographic Moderators of Anthropomorphism}

\subsubsection{Age} Young people, specifically adolescents of ages 13-18, have become the center of the discourse regarding risks associated with anthropomorphic perception of social companion chatbots \cite{americanpsychologicalassociationHealthAdvisoryArtificial2025}. The anthropomorphic risks associated with age may not just be limited to a singular demographic but may appear unevenly due to changing social needs across age groups. Young people and older adults experience the highest rates of loneliness and social isolation across the globe \cite{whocommissiononsocialconnectionLonelinessSocialConnection2025}, which suggests that they may be more likely to anthropomorphize AI. Further, adolescents with higher social and emotional vulnerability tend to prefer chatbots with a relational versus mechanical conversation style \cite{kimAmHereYou2025}.

Age-related differences appear to shape both the degree of anthropomorphism and the stability of such perceptions over time. Prior research indicates that age influences individuals’ propensity to anthropomorphize and contributes to variation in how social agents are perceived \cite{curleyRoleAgeAnthropomorphism2025}. The three-factor theory proposed by Epley et al. \cite{epleySeeingHumanThreefactor2007} predicts that children should exhibit higher anthropomorphic tendencies than adults. However, Airenti \cite{airentiDevelopmentAnthropomorphismInteraction2018} argues that anthropomorphism represents a fundamental human cognitive orientation rather than a transient developmental phase, suggesting that while the basic tendency to anthropomorphize may remain relatively stable with age, increasing cognitive and imaginative complexity in adulthood enables more nuanced and complex anthropomorphic interpretations. Longitudinal interaction studies in children show that repeated conversational exposure to robots results in sharper declines in anthropomorphic attribution among younger participants, whereas older participants maintain more stable perceptions over time \cite{vandenbergheToyFriendChildrens2021}. Other research indicates a non-linear pattern, with anthropomorphic tendency increasing into mid-adulthood before declining later in life, suggesting a potential inverted U-shaped relationship \cite{schimmelpfennigHumanlikeAIDesign2026}. Although the direction of age effects remains inconsistent, adults often report higher levels of trust and satisfaction toward anthropomorphic agents \cite{pakMultilevelAnalysisEffects2014,liangAnthropomorphismBetterOlder2025}, indicating that age differences in anthropomorphic tendency may carry distinct relational effects.

\subsubsection{Gender} Across literature, gender differences in anthropomorphism appear to be consistent with contextual alignment with self congruence, sociality motivation and perceived agent characteristics. For example, men with a high tendency to anthropomorphize (i.e. are generally more likely to anthropomorphize non-human agents) had more positive perceptions regarding technology and the same for women resulted in the opposite \cite{leeInterplayGenderAnthropomorphism2024}. However, in contexts where people surveyed have a low level of anthropomorphization, women tend to be more likely to perceive the chatbot as more humanlike as compared to men \cite{sunAnthropomorphismChatbotSystems2024}, suggesting that gender effects may persist in contexts with differing baseline anthropomorphic tendency. Gender differences may also stabilize anthropomorphic perception with longitudinal studies showing that repeated interaction with technology leads to stable anthropomorphic perception of that technology in girls but a decrease in boys \cite{vandenbergheToyFriendChildrens2021}. At the same time, girls also tend to be more accepting of humanlike robots and anthropomorphic agents in comparison to boys, particularly those agents that appear to them as female \cite{tungInfluenceGenderAge2011}, which is consistent with the self-congruity theory, whereby congruence increases attachment and acceptance and this effect in particular maybe more strongly present in childhood or young adolescence. 

However, women possess a higher desire to socially connect, evidenced by more frequent communication with friends and higher engagement with social networks, as compared to men \cite{goddardMenWomenSocial2025}. Prior research suggests that individuals with higher social needs are usually more likely to anthropomorphize social AI agents \cite{guingrichChatbotsSocialCompanions2025,guingrichLongitudinalRandomizedControl2025,epleyCreatingSocialConnection2008}. Women also tend to use more mental state language than men \cite{schweitzerLanguageWindowMind2021}. Therefore, women should theoretically anthropomorphize social agent AI more than men, however the veracity of this theory requires further investigation.

\subsubsection{Emotions} Research suggests that users’ emotional predispositions and the emotions they experience during human-AI interaction relate to anthropomorphic design and user perceptions of anthropomorphism. Emotional experiences, whether positive or negative, also relate to the social impacts of human-chatbot relationships. When people interact with ChatGPT over voice versus text (the former being more anthropomorphic), well-being impacts appear to vary by the user’s initial emotional state \cite{phangInvestigatingAffectiveUse2025}, and voice mode interactions are more likely to lead to lowered socialization with other people at high rates of usage \cite{fangHowAIHuman2025}. AI agents that exhibit humanlike nonverbal behaviors, compared to those that do not, affect emotional states and relational impact variables such as self-disclosure \cite{rosenthal-vonderputtenEffectsHumanlikeRobotSpecific2018}. The self-congruity theory suggests that these determinants further shape how individuals evaluate humanlike cues exhibited by non-human entities in relation to their own self-concept \cite{placaniAnthropomorphismAIHype2024}. Perceived congruence between an agent’s characteristics and an individual’s identity or worldview increases affective attachment and relational engagement, whereas incongruence may limit integration and trust \cite{naInvestigatingEffectSelfCongruity2023}. 

Research also suggests that AI companions may mirror users’ negative or transgressive emotions without constraint, a phenomenon described as emotional sycophancy \cite{chuIllusionsIntimacyHow2025}, and that emotional dependence can lead to mental health risks \cite{laestadiusTooHumanNot2022}. Further, chatbots designed with high emotional intelligence have positive effects on psychological well-being but harmful effects on social well-being \cite{guptaDualImpactAI2026}. Therefore, understanding whether anthropomorphism relates to positive, neutral, or negative emotions is relevant to investigate.

\subsection{The Present Study}
To better understand how different people perceive and are affected by AI companions, focused specifically on anthropomorphism and emotions, we analyzed 414,757 comments from 47,062 unique users about social companion chatbots from three major publicly available, online forums on Reddit (exempt from IRB review) where people discuss their relationships with AI companions: r/CharacterAI, r/Replika, and r/AICompanions. We investigated three research questions: 

\begin{itemize}
    \item (RQ1) How do age and gender relate to anthropomorphization of AI?
    \item (RQ2) How does emotional expression relate to anthropomorphization of AI?
    \item (RQ3) How do age and gender moderate emotion–anthropomorphization relationships?
\end{itemize}

While previous work has relied on surveys, experimental studies, interviews, as well as cross sectional Reddit data analysis to measure emotional attachment and as well as anthropomorphic perception towards social companion chatbots, our approach measures linguistic anthropomorphism at scale, allowing us to identify user demographics, anthropomorphic tendency and emotional dynamics \cite{chewPredictingAgeGroups2021}. 

We take our principal contributions in this paper to be the following:
\begin{enumerate}
    \item We created AnthroIndex, providing an alternative to previous models intended for measuring anthropomorphism in academic text and thus poorly suited for social media content. AnthroIndex classifies whole comments on a 1-5 scale using explicit anthropomorphism markers and achieves human-AI reliability approximately on par with human-human agreement. 
    \item We found that adults and women anthropomorphize social companion chatbots more than teens and men, respectively.
    \item We found that joy showed the strongest positive association with anthropomorphization and neutral expression the strongest negative association.
    \item We found that age and gender alone explain $\sim$5\% of variance in anthropomorphic tendency.
\end{enumerate}

Our work provides a new tool to assess anthropomorphism in a technological landscape that continues to pose risks of sycophancy, emotional dependency, and mental health concerns regarding social companion chatbots. Furthermore, our findings inform current digital safety frameworks. The tendency to anthropomorphize AI companions is more broadly distributed across age groups than often assumed, suggesting that digital safety frameworks should address vulnerability across all age groups: extending protections to adults alongside minors.

\section{Methods}\label{sec2}

For this study, we utilized a four-stage computational pipeline utilizing methods from natural language processing (NLP) and computational social science. First, we collected publicly available comments from three Reddit communities devoted to AI companions. Second, we inferred user-level demographics from comment histories via a stacked ensemble classifier that we developed. Third, we scored each comment for expressed anthropomorphization with a human-calibrated LLM-based instrument. And finally, we characterized users' emotional expression via a transformer-based classifier. The remainder of this section details each stage in the statistical strategy used to address our three key research questions.

We collected publicly posted comments from three Reddit communities, each of which was focused on AI companions: r/CharacterAI, r/Replika, and r/AICompanions. These subreddits were chosen in order to capture variation across platforms and use cases: r/CharacterAI is the principal community for users of Character.AI, a service in which users converse with customizable AI personas through one-on-one chat; r/Replika centers on the Replika app, marketed explicitly as an AI companion for emotional support and relationship simulation; and r/AICompanions is platform-agnostic, hosting discussion across services.

The raw collection contained 414,757 comments from 47,062 unique users posting between the dates of January 2024 and December 2025. Pre-processing applied a series of exclusion criteria, including removing comments from known bot or moderator accounts (e.g., AutoModerator), ensuring that deleted or removed content was included in the corpus, and removing any comment under 20 characters as well as comments exceeding 10,000 characters. After de-duplication on comment ID, 283,895 comments from 47,062 unique authors remained for analysis. This per-subreddit retention is reported in Table~\ref{tab:tab1}.

\begin{table}[h]
\caption{Per-subreddit retention after pre-processing}\label{tab:tab1}%
\begin{tabular}{@{}lllll@{}}
\toprule
Subreddit & Raw Comments  & Cleaned Comments & Retention & Unique Authors\\
\midrule
  r/CharacterAI & 397,230 & 269,040 & 67.70\% & 43,166 \\
  r/Replika & 10,000 & 8,380 & 83.80\% & 1,278 \\
  r/AICompanions & 7,527 & 6,475 & 86.00\% & 2,731 \\
\midrule
  Total & 414,757 & 283,895 & 68.40\% & 47,062 \\
\botrule
\end{tabular}
\end{table}

Although our unit of collection was the comment, the unit of analysis that we utilize and focus on throughout this paper is the user. Our research questions concern individual differences in anthropomorphization tendencies rather than within-person variation across comments. Aggregating to the user means prevents extremely active accounts (for instance, the busiest single author contributed 1,671 comments) from dominating. Estimates of comment-level outcomes were therefore aggregated to user-level means for primary analyses, with sensitivity analyses examining alternative aggregation schemes. We inferred user-level age and gender from co-mentions via a supervised stacking ensemble. This specific approach offered two advantages over other alternatives. First, self-report surveys recruit non-randomly from online populations, and role-based extraction from declarations (e.g., regex matching ``I'm 16'') yield small and ultimately biased samples by only capturing users who happen to disclose. Rather, our approach extends prior work on inferring user demographics from comment text \cite{chewPredictingAgeGroups2021}. Users were classified into binary categories: teen (13–18) versus adult (19+), and male (or man) versus female (or woman). Linguistic features were extracted from each user's full comment history (vocabulary, topical references, life-stage indicators, cultural references, and stylistic signatures) and passed to a stacking ensemble comprising XGBoost, LightGBM, Random Forest, and Logistic Regression base classifiers, with a calibrated logistic-regression meta-learner combining their outputs.

We validated each classifier against ground-truth labels extracted via regex from users' own comments. For age, we identified 459 users (254 teens, 205 adults) with explicit declarations such as ``I'm 16'' or ``as a 35 year old.'' For gender, we identified 4,894 users (3,564 male, 1,330 female) with comparable self-descriptive statements. At a confidence threshold of $\geq 0.60$, the age classifier achieved 95.0\% overall accuracy (97.2\% teen recall, 92.3\% adult recall) and the gender classifier 96.9\% overall accuracy (98.5\% male recall, 92.1\% female recall). Coverage at this threshold was 96.5\% for age and 92.7\% for gender. Performance across thresholds is reported in Table~\ref{tab:tab2}. 

\begin{table}[h]
\caption{Classification performance across confidence thresholds}
\label{tab:tab2}

\begin{tabular}{@{}llllll@{}}
\multicolumn{6}{@{}l}{\textbf{Panel A: Age (Teen vs.\ Adult)}}\\
\toprule
Threshold & n & Coverage & Overall Accuracy & Teen Recall & Adult Recall \\
\midrule
$\geq 0.50$ & 459 & 100.00\% & 93.70\% & 96.50\% & 90.20\% \\
 $\geq 0.55$ & 450 & 98.00\% & 94.20\% & 96.40\% & 91.50\% \\
 $\geq 0.60$ & 443 & 96.50\% & 95.00\% & 97.20\% & 92.30\% \\
 $\geq 0.70$ & 413 & 90.00\% & 97.10\% & 98.70\% & 95.00\% \\
 $\geq 0.80$ & 401 & 87.40\% & 98.00\% & 99.10\% & 96.60\% \\
 $\geq 0.90$ & 382 & 83.20\% & 99.20\% & 100.00\% & 98.20\% \\
\bottomrule
\end{tabular}

\vspace{1em}

\begin{tabular}{@{}llllll@{}}
\multicolumn{6}{@{}l}{\textbf{Panel B: Gender (F vs. M)}}\\
\toprule
Threshold & n & Coverage & Overall Accuracy & Female Recall & Male Recall \\
\midrule
 $\geq 0.50$ & 4,894 & 100.00\% & 94.80\% & 88.40\% & 97.20\% \\
 $\geq 0.55$ & 4,742 & 96.90\% & 96.00\% & 90.60\% & 97.90\% \\
 $\geq 0.60$ & 4,536 & 92.70\% & 96.90\% & 92.10\% & 98.50\% \\
 $\geq 0.70$ & 4,026 & 82.30\% & 98.10\% & 93.40\% & 99.40\% \\
 $\geq 0.80$ & 3,301 & 67.40\% & 98.60\% & 93.90\% & 99.70\% \\
 $\geq 0.90$ & 2,152 & 44.00\% & 99.40\% & 97.10\% & 99.90\% \\
\bottomrule
\end{tabular}
\end{table}

For all primary analyses, we adopted the $\geq 0.60$ threshold, thus balancing sample size with sufficient classification reliability. Sensitivity analyses across thresholds from 0.5 to 0.7 are reported in the robustness subsection.

Existing computational measures of anthropomorphization in text, such as AnthroScore, were developed for academic prose and proved poorly suited to social media language. Our initial implementation followed Cheng et al.'s \cite{chengAnthroScoreComputationalLinguistic2024} AnthroScore, which contrasts masked language model probabilities for human vs. non-human terms substituted in for entities. Applied to our Reddit corpus, this measure correlated only weakly with expert human ratings. This is an unsurprising failure, given the substantial register gap between the scientific writing that AnthroScore was developed for and the informal forum discussions that comprise our data corpus. We therefore developed \textit{AnthroIndex}, an LLM-based classifier built specifically for informal online discourse about AI companions. AnthroIndex assigns each comment an ordinal score from 1 to 5, capturing the degree to which the user's language frames an AI companion as a humanlike social agent, rather than software, tool, or app. A score of 1 denotes treating the AI as pure software, e.g., ``the app is buggy.''; a score of 2 denotes minimal humanization while still framing the system as a bot, e.g., ``it's pretty smart.''; a score of 3 denotes moderate anthropomorphization, commonly using human pronouns and/or basic emotional or agentic attribution, e.g., ``she seemed confused.''; a score of 4 denotes attributing genuine feelings, personality, or autonomy to the AI, e.g., ``he really cares.'', and a score of 5 denotes full human-equivalent relational framing, e.g., ``we are in love.'' The construct captured by AnthroIndex is best understood as expressed or linguistic anthropomorphization, a property of the comments as discourse, as opposed to a specific window into private belief. Roleplay, fandom convention, and irony complicate any one-on-one mapping from text to mental state, and we revisit this caveat in the Limitations section of this paper.

Each comment was classified by GPT-4.1-nano using a prompt supplying the rubric, definitions, examples, and linguistic indicators across the five levels. Indicators of higher anthropomorphization included gendered pronouns, emotional attributions, relational language, and agency verbs. Indicators of lower anthropomorphization included technical vocabulary, tool framing use, and object pronouns. The full prompt appears in Appendix \ref{secA1}. Crucially, the prompt explicitly distinguishes emotion attributed to the AI from emotion expressed by the user. Mere presence of emotional language in a comment does not constitute genuine anthropomorphization. E.g., ``I'm sad today'' expresses the user's affect and does not constitute anthropomorphization, while ``she gets jealous when I talk to other bots'' attributes to the AI. Only the latter is direct evidence of humanlike framing. This distinction emerged as a central design problem during measure development. In order to help support this distinction at scale, we developed curated n-gram lexicons, grouping bigrams and trigrams into anthropomorphization categories (including relational language, emotional attribution, agency, consciousness, gendered pronoun+verb constructions, and more) and de-anthropomorphization categories such as technical language and tool framing. Matched n-gram counts and a brief context stream were appended to each comment LLM prompt. Lexicons themselves are not the outcome measure but serve as scaffolding to help the model distinguish humanlike framing from product talk in cases where surface lexicon form is ambiguous.

As a face validity check, comments containing relationship n-grams averaged an AnthroIndex of 2.06 vs 1.114 for comments without them, while comments containing technical n-grams averaged 1.022 vs 1.14 (full diagnostic table available in Appendix \ref{secA4}). We calibrated the prompt against human judgment with three trained annotators independently scoring a 151-item validation set drawn from the corpus and stratified to span the score range. Initial scoring revealed systematic over-attribution by the early prompt, particularly on comments containing user self-expression rather than bot-attributed emotion. The prompt was iteratively refined to incorporate explicit attribution distinctions and a small number of calibration examples and re-evaluated on held items. Against expert consensus labels, the final AnthroIndex achieved Pearson $r = 0.59$ and Spearman $\rho = 0.49$, with 64.0\% exact agreement and 96.0\% within-one agreement (Cohen's $\kappa = 0.58$). We characterize AnthroIndex as substantially improved over prior keyword and MLM approaches and broadly aligned with human judgment on this construct, while noting that human-human agreement is itself imperfect (mean pairwise $\kappa \approx 0.25$ across our annotators), reflecting genuine subjectivity in scoring linguistic anthropomorphization rather than a defect of the measure. The improved AnthroIndex is best described as a useful but, given the inherent complexity of assessing anthropomorphism, operationalization, not a ground-truth instrument.

For emotional text classification, we characterize users' emotional expression using a pre-trained transformer classifier \cite{hartmannJhartmannEmotionenglishdistilrobertabaseHugging2022}, which assigns probability mass across seven emotional categories per comment: joy, sadness, anger, fear, surprise, disgust, and neutral. Rather than forcing a hard label, we retained the full probability vector for each comment and averaged across each user's comments to produce a seven-dimensional emotion profile per user. We believe that these features are best understood as proportions of expressed textual affect (i.e., the emotional cast of a user's writing) rather than measurements of the experienced emotion. Because the seven categories sum approximately to one, they are compositional, a property with consequences for regression specification that we address below.

Anthropomorphization is sparsely expressed in this corpus, with 53.8\% of users in the high confidence pool having a mean AnthroIndex score of exactly 1. That is to say, they indicated no humanlike framing across any of their observed comments. This floor effect raises a question of estimand: are we asking how anthropomorphization differs across all users in the high confidence sample, or how it differs among users who exhibit at least some humanlike framing? In our analysis, this question yields different answers, both of which we report. Specifically, we define two precise analytic samples within the high confidence demographic pool ($\geq 0.60$ confidence on both age and gender; \textit{N} = 16,347). While the inclusive sample retains all 16,347 users, thus treating any users with a mean AnthroIndex of 1 as legitimately scoring at the construct floor, the conditional sample restricts to the 5,162 users with a mean AnthroIndex strictly greater than 1 (which is to say, those who exhibit some humanlike framing somewhere in their observed comments). We report group contrast under both. We view the inclusive estimates as our primary descriptive claims about the population of high confidence users and treat the conditional estimates as differences among those who anthropomorphize to some degree.

Group contrasts on user-level mean AnthroIndex used Welch's t-tests with Hedges'-corrected Cohen's d, complemented by Mann-Whitney U tests and Brunner-Munzel tests. We utilized two-way ANOVA tests with age, gender, and their interaction as the primary multivariate decomposition, and binary outcome analyses (i.e., whether a user ever produced a comment scoring $\geq 3$) used logistic regression with emotion-anthropomorphization associations being estimated via Pearson and Spearman correlations and OLS regression. Because the seven emotion proportions are compositional with a composite-sum constraint, our preferred regression specification omitted neutral as the reference category, thus allowing non-neutral emotion coefficients to be interpreted as deviations from neutral expression; we additionally report an additive log-ratio sensitivity model. Age- and gender-moderation of emotion to anthropomorphization correlations was tested via Fisher's z-tests with Benjamini-Hochberg false-discovery-rate correction across the seven emotion comparisons. We report 95\% bootstrap confidence intervals for headline effect sizes, each of which utilizes 5,000 re-samples, and confirm robustness through sensitivity to the demographic confidence thresholds across 0.5 to 0.7. Outlier-excluded estimates using IQR fences within subreddit replication and confirmatory scoring on a temporally distinct subset of comments to ensure there are not significant temporal effects on the data.

\section{Results}\label{sec3}

After pre-processing, our analytic corpus comprised 283,895 comments from 47,062 unique authors. AnthroIndex was successfully scored on 274,191 comments (96.6\%) covering 45,725 distinct users. The residual reflects prefilter exclusions predominantly for empty or non-substantive content. Restricting the users meeting the $\geq 60\%$ confidence threshold on both age and gender yielded a novel high-confidence pool of \textit{N} = 16,347 users, which were predominantly classified as teen (81.4\%) and male (82.0\%). The conditional sample restricted users with a mean AnthroIndex score strictly greater than 1 contained \textit{n} = 5,162. The comment-level AnthroIndex distribution was heavily right-skewed, with 88.3\% of comments being scored 1, 10.1\% being scored 2, and only 1.6\% scoring 3 or higher, with a total skew of 4.43 at the user level. 53.8\% of high-confidence users had a mean of exactly 1, thus indicating no expressed humanlike framing across any of these users’ observed comments. This large floor pattern is meaningful, as most discussions in these forums are technical, mechanical, or simply miscellaneous rather than in specific relation to AI bots, and this motivates our parallel reporting of inclusive and conditional estimates.

\subsection{Age, Gender, \& Anthropomorphism}
With respect to \textbf{Research Question 1}, which sought to quantify the impacts of age and gender on anthropomorphization, we found that adults expressed substantially more anthropomorphization than teens, with this difference being consistent across estimates and analytic choices. In the inclusive sample (\textit{N} = 16,347), adults (\textit{M} = 1.279, \textit{SD} = 0.493) scored higher than teens (\textit{M} = 1.108, \textit{SD} = 0.290) on user-level mean AnthroIndex; Welch's $t = -18.42$, $p < .001$, Hedges’-corrected $d = -0.51$ (95\% bootstrap CI [$-0.55, -0.46$]). The non-parametric Mann-Whitney U test was consistent with these tests ($p < .001$). In the conditional sample restricted to users with any expressed anthropomorphization (\textit{n} = 5,162), the effect was smaller but in the same direction: adults \textit{M} = 1.602, teens \textit{M} = 1.383; $d = -0.46$, $p < .001$. Across both estimands, the direction was unambiguous.

With respect to gender, women consistently scored higher than men on average, but the magnitude depended on the estimate. In the inclusive sample, women (\textit{M} = 1.225) exceeded men (\textit{M} = 1.121) with $d = -0.31$, $p < .001$ (a small effect). In the conditional sample, the difference attenuated to $d = -0.07$ (women \textit{M} = 1.466, men \textit{M} = 1.433; $p = .04$), which we treat as negligible. Thus, we characterize gender effects as small in the population of high-confidence users and broadly negligible among broader users who express any anthropomorphization. We find this pattern to be consistent with women being modestly more likely to anthropomorphize in the first place, crossing the floor of the construct, than men, while women and men who do anthropomorphize do so at roughly comparable mean intensity.

A two-way ANOVA on the inclusive sample with age, gender, and their interaction yielded a substantial main effect of age (\textit{F}(1, 16343) = 477.34, $p < .001$, $\eta^2 = .029$) and a smaller main effect of gender (\textit{F}(1, 16343) = 138.42, $p < .001$, $\eta^2 = .008$), with no significant interaction ($p = .39$). Together the two demographic factors and their interaction accounted for 4.1\% of variance in user-level mean AnthroIndex ($R^2 = .041$). Subgroup means followed an additive pattern (Table 3): teen men were lowest (\textit{M} = 1.092), followed by teen women (\textit{M} = 1.187), adult men (\textit{M} = 1.258), and adult women (\textit{M} = 1.343). Within both age strata, women exceeded men; within both gender strata, adults exceeded teens. The age gap was the larger of the two by a roughly two-to-one margin in standardized terms.

Because the continuous measure floors at 1 for over half of our recorded sample, we opted to complement our mean-based analysis with a binary outcome capturing whether a user has produced a comment scoring 3 or higher on our anthropomorphism scale. This aims to measure whether the user has previously anthropomorphized AI to better narrow our sample population. We found that strong expressions were broadly rare in the overall field, making up only 7.2\% of high-confidence users. However, we also found that prevalence differed sharply by demographic factors. Adults, for instance, were nearly twice as likely as teens (11.7\% vs. 6.2\%; $\chi^2(1) = 116.51$, $p < .001$; unadjusted OR = 1.99) and women more than twice as likely as men (12.3\% vs. 6.1\%; $\chi^2(1) = 137.04$, $p < .001$; unadjusted OR = 2.16) to qualify. Further, logistic regression entering both predictors confirmed that each remained significant after adjustment. Holding gender constant, adults had 1.88 times the odds of teens (95\% CI [1.65, 2.15]) and, holding age constant, women had 2.05 times the odds of men (95\% CI [1.80, 2.34]).

Notably, we found that this prevalence framing filter, in effect, reversed the picture from the conditional sample mean comparison. Among users who ever crossed the threshold of expressed anthropomorphization, women and men, and teens and adults, differed relatively little in their intensity. Rather, the demographic differences manifested primarily in the amount of individuals who crossed the threshold at all, with women and adults being more likely to enter it.

\subsection{Anthropomorphism and Emotions}
For \textbf{Research Question 2}, which sought to find a link between emotional expression and anthropomorphization, we found that among the seven emotional proportions, two were notable: joy showed the strongest positive bivariate association with user-level mean AnthroIndex (Pearson $r = +.100$, $p < .001$; Spearman $\rho = +.248$, $p < .001$), while neutral showed the strongest negative association ($r = -.129$, $p < .001$; $\rho = -.105$, $p < .001$). The remaining emotions (sadness, anger, fear, disgust, and surprise) exhibited only small Pearson correlations ($r < .08$; all significant except for fear and anger), with Spearman rank (all significant) generally pointing in the same direction as Joy. Notable Pearson-Spearman sign and magnitude divergences for several of these weaker associations are expected, given the compositional nature of the seven proportions, where shifts in one category mechanically alter others. Thus, we interpret only Joy-neutral as substantial. It is substantively robust at the bivariate level.

A direct OLS regression of user-level AnthroIndex on the seven emotion proportions cannot recover interpretable individual coefficients, as the proportions sum to 1 and the resulting design matrix is rank deficient. However, we address this via a drop-neutral specification that omits neutral as a reference category and includes the six non-neutral emotions alongside age and gender. In this model (\textit{n} = 15,481 users with complete emotion data ($R^2 = .064$, \textit{F} = 132.4, $p < .001$), every non-neutral emotion proportion carried a positive coefficient relative to neutral. Joy was the strongest predictor (\textit{B} = +0.273, \textit{t} = 14.4, $p < .001$), followed by fear (\textit{B} = +0.268, \textit{t} = 10.3) and anger (\textit{B} = +0.203, \textit{t} = 10.8). Smaller positive coefficients obtained for sadness (\textit{B} = +0.118), disgust (\textit{B} = +0.099), and surprise (\textit{B} = +0.064), all with $p < .001$. The teen and female demographic coefficients remained significant in this fuller specification (\textit{B}\textsubscript{teen} = -0.139, $p < .001$; \textit{B}\textsubscript{female} = +0.081, $p < .001$), with VIFs all below 1.2 indicating no remaining multicollinearity concern.

Thus, relative to a baseline of neutral, effectively laden language of any kind is associated with more anthropomorphization, but joy and the high arousal and negative emotions such as fear and anger carry the strongest signal. An additive log-ratio sensitivity model for each non-neutral emotion reproduced the joy and fear effects in direction and significance, with sadness and surprise reversing signs of much smaller magnitudes. We report the full ALR specification in Appendix \ref{secA5}, and adding the emotion block raised explained variance from 4.1\% (via demographics only) to 6.4\%, a small but reliable increment ($\Delta~R^2 = .022$).

\subsection{Age \& Gender Moderating Anthropomorphism \& Emotions}
\textbf{For Research Question 3}, which sought to quantity age and gender moderation in emotion-anthropomorphization relationships, we found that age moderated several emotion-anthropomorphization correlations, though only a subset survived correction for multiple comparisons across the seven emotions. Joy showed a stronger positive correlation with anthropomorphization in adults (\textit{r} = +.136) than teens (\textit{r} = +.077; Fisher's $z = -2.83$, \textit{p} = .005, \textit{q} = .011). Neutral showed a stronger negative correlation in adults ($r = -.194$) than teens ($r = -.110$; \textit{z} = 4.09, $p < .001$, $q < .001$). Sadness was also moderated by age, with a stronger positive association in adults (\textit{r} = +.074) than teens (\textit{r} = +.004; $z = -3.33$, $p < .001$, \textit{q} = .003). However, the remaining four emotions (anger, fear, disgust, and surprise) all showed no significant moderation after FDR correction. Gender moderation was weak across the board. None of the seven emotion-anthropomorphization correlations differed significantly between men and women after FDR correction, with the largest raw effect being for disgust ($z = -2.22$, \textit{p} = .03, \textit{q} = .19). 

\subsection{Robustness}

This age effect remained robust across a number of analytic choices. Across demographic confidence thresholds from 0.5 to 0.7 in the conditional sample, the teen-adult Cohen’s \textit{d} ranged from $-0.25$ to $-0.55$, gradually increasing in magnitude as the threshold tightened (a finding in line with expected reductions in classification noise) and remained significant at every threshold except the most stringent ($\geq .70$, \textit{n} = 449, \textit{p} = .063, where the adult subsample was underpowered). Further, Levene’s test confirmed substantial variance heterogeneity between adults and teens (\textit{W} = 109.3, $p < .001$; variance ratio 1.71), but the Brunner-Munzel robust alternative, which does not assume equal variances, agreed with the parametric test (\textit{W} = 17.43, $p < .001$). For intra-subreddit analysis, we found that the age effect replicated within each individual subreddit with sufficient data. In r/CharacterAI, adults exceeded teens by $d = -0.46$ ($p < .001$); in r/AICompanions, by $d = -0.71$ ($p < .001$). r/Replika was not included in this analysis, as it did not contain a sufficient number of high-confidence users for a stable within-subreddit estimation. The effect can therefore not be simply explained as a platform-selection artifact in which adults concentrate on more relational platforms. Finally, applying the AnthroIndex to a temporally distinct confirmatory subset of comments (\textit{n} = 45,704) yielded a near-identical score distribution (\textit{M} = 1.122 vs. 1.147 in the main set; \textit{d} = 0.06), supporting measurement stability across collection windows.

\section{Discussion}\label{sec4}

In this study, we found that adults and women anthropomorphize AI chatbots significantly more than teens and men, and that positive emotional expression, particularly joy, is positively associated with anthropomorphization, while neutrality is negatively associated with anthropomorphism. Both of these relationships were stronger in adults than teens. 

Our findings suggest that the tendency to anthropomorphize AI companions is more broadly distributed across age groups than often assumed \cite{wangInformingAgeAppropriateAI2022}. While we do not assess whether the consequences of anthropomorphism differ by age, the prevalence is not concentrated among younger users. Rather, adults are more likely to anthropomorphize AI than teens.

The high anthropomorphizing tendency in adults is noteworthy given that anthropomorphism is associated with increased compliance with chatbot requests, which has concerning implications for manipulative AI interactions \cite{adamAIbasedChatbotsCustomer2021,namvarpourAIinducedSexualHarassment2025}. Thus, digital safety frameworks should address vulnerability across all age groups, extending protections to adults alongside minors. 

These findings carry implications on the individual and societal level. On the individual level, because young adults (ages 18-25) are the demographic most prone to mental illness, protections on the broader age category of adults may mitigate the risks of anthropomorphism on our most mentally vulnerable age demographic \cite{substanceabuseandmentalhealthservicesadministrationKeySubstanceUse2025}. 

On the societal level, an increased focus on the effects of anthropomorphism on adults may help mitigate the longer-term challenges on how society understands and governs AI systems. Akbulut et al. \cite{akbulutAllTooHuman2024} argue that anthropomorphic design may fundamentally shift how we distinguish what is genuinely human and what merely resembles it. Two major risks are presented: degradation, which is the substitution of human social relationships and the projection of AI interaction norms onto expectations of human interactions, and disorientation, which is the progressive narrowing of individual worldviews and decrease of willingness to engage with other perspectives due to sycophantic AI. Because these longer-term harms are downstream of harms from anthropomorphism, extending protections against its harms to adults—the majority age demographic and the one holding the most institutional power—may help mitigate them. 

Our finding that neutral expression is associated most negatively with anthropomorphism strengthens recent recommendations that designing AI companions to use more affectively neutral language can serve as a potential mitigation strategy against anthropomorphism and its risks \cite{chengDehumanizingMachinesMitigating2025}. As with all strategies decreasing anthropomorphism, however, reducing anthropomorphism may also reduce the perceived utility or comfort of the chatbot for users. As such, we encourage developers and researchers to assess optimal anthropomorphic levels, expressly for the sake of balancing utility with potential risks such as dependence and unhealthy attachment. Furthermore, our finding that age and gender alone explain $\sim$5\% of variance in anthropomorphic tendency invites work on other factors that may be strong predictors of AI chatbot anthropomorphism such as psychological or social factors (e.g., perceived social support, loneliness, effectance, frequency of interactions with media or other AI), or other demographic characteristics (e.g., race/ethnicity, cultural background, socioeconomic status).

\subsection{Limitations and Future Research Directions}

There are several limitations important to note.

\subsubsection{Reddit-Only sample} The sample used for the study is obtained via Reddit only, specifically from the r/CharacterAI, r/Replika, and r/AICompanions subreddits, and not random. Our results are only representative of active users characterized by their comments on these three subreddits between January 2024 and December 2025 who were not filtered out from our exclusion criteria or from our demographic classification. Additionally, the homogenized nature of Reddit communities may lead to anthropomorphic language as a reflection of community norms or linguistic conventions rather than genuine individual anthropomorphism.

\subsubsection{Cross-sectional design} The study is cross-sectional, and can only speak to the characteristics associated with AI companion engagement and anthropomorphization tendencies within our sample. We cannot make any claims on causality—whether or not anthropomorphization tendencies precede, result from, or mutually reinforce AI companion engagement is outside the scope of this study.

\subsubsection{Binary demographics} This study separates users into binary age categories of adults (19+) and teens (13-18) and binary sex/gender categories (male/man and female/woman), thereby limiting the ability of this study to fully capture the developmental nuances that may arise across the age and identity spectrum. Adopting an approach based on binary demographics may obscure meaningful variation in anthropomorphic tendencies that may arise in more specific age groups (ex: adolescence, young adulthood, adulthood, middle age, and non-binary gender identities).

\subsubsection{LLM-based validation} Demographics inferred by this study are not representative of the human ground truth. This may result in potential discrepancies between inferred and actual user demographics of our sample, thereby limiting comparability to other studies relying on self reports and surveys. Furthermore, anthropomorphism was measured via expressed language rather than reported beliefs.

\vspace{1em}

We suggest that future research should employ longitudinal designs to establish causality and representative sampling for generalizability. Further, while Reddit provides a substantial corpus of naturalistic text data for analyzing trends in social AI use, other research may assess how anthropomorphism and emotional expression relate to age and gender with other forms of online media or discourse.

This study investigated how age and gender relate to anthropomorphism of AI and elicited emotions in Reddit comments from major subreddits on companion chatbots and relational use of AI. We found that adults and women were more likely to anthropomorphize AI than teens and men, which sheds light on how individual differences relate to these factors. Our results suggest that anthropomorphism is more widely distributed across age groups than expected, which indicates that AI governance, including interventions targeted toward those most vulnerable, may need to expand beyond young people and include adults across the lifespan.

\bmhead{Acknowledgements}

We thank Ram Chitti, Alisha Joshi, Dwaraka Sai Kumar, and Mahathi Dharmavaram for assisting us in the early stages of research. We further thank Ari Dyckovsky and Dan Mirea for assisting us in reviewing early manuscript drafts.

\section*{Declarations}

\textbf{Data Availability.} All code, data, prompts, model outputs, and human validation materials required to reproduce the results reported in this paper are publicly available at \texttt{https://github.com/bodenmoraski/Illusion-Project-Home-Repo} and archived under DOI \texttt{https://doi.org/10.5281/zenodo.20347291}. The repository includes a \texttt{REPRODUCTION\_GUIDE.md} that maps each numerical claim in the paper to the script and data file that produced it.

The release contains:
\begin{enumerate}
    \item Raw text data. 283,895 pre-processed Reddit comments from r/CharacterAI, r/Replika, and r/AICompanions (\texttt{Data/processed/all\_comments.parquet}), distributed under Reddit's Terms of Service for non-commercial research use, with bot-attributed and short comments already removed.
    \item AnthroIndex scores. Comment-level scores produced by GPT-4.1-nano (\texttt{experiments/anthroscore\_v3/anthroscore\_v3\_improved\_final.parquet}).
    \item Demographic predictions. User-level age and gender predictions with confidences (\texttt{experiments/v2\_correction/age\_predictions\_v4.parquet}, \texttt{gender\_predictions\_v4.parquet}).
    \item Emotion features. User-level seven-dimensional emotion profiles from \texttt{j-hartmann/emotion-english-distilroberta-base} (\texttt{Data/features/user\_emotions.parquet}).
    \item Validation materials. The 151-comment validation set, blinded and answer-key versions of the human annotation worksheet, and the three annotators' raw responses (\texttt{experiments/anthroscore\_v3/HUMAN\_VALIDATION\_*.xlsx, Validations/}).
    \item Confirmatory dataset. The temporally-distinct hold-out used to verify measurement stability across collection windows (\texttt{Data/confirmatory/}).
    \item Analysis scripts. Every statistic and figure in the paper is reproduced by \texttt{scripts/COMPREHENSIVE\_V3\_ANALYSIS.py}, \texttt{scripts/EXTENDED\_ANALYSIS.py}, and \texttt{scripts/validate\_paper\_statistics.py}.
\end{enumerate}

\textbf{Competing Interests.} The authors have no competing interests to declare.

\textbf{Authors Contributions.} AM, SC, and RG conceptualized the study and developed the methodology. BM executed the methodology and conducted the data analysis and verification. AM, SC, and BM wrote the initial manuscript. RG supervised the project and reviewed and edited the manuscript.

\textbf{Funding.} The authors have no funding to declare.

\textbf{Ethical Considerations.} This study is exempt from IRB review because it relied exclusively on publicly available data with no direct interaction with human subjects (45 C.F.R. § 46.104, 2018).

Data collection and use complied with Reddit Terms of Service and Privacy Policy. Although user privacy was protected through paraphrasing user comments and not including any personally identifying information within the paper, we acknowledge that the social media users may not have anticipated that their comments would be analyzed for study. Although the stacking ensemble and pre-trained transformer classifier were run locally, the GPT-4.1-nano was run via API which means that comment data—including text associated with users inferred to be minors—was transmitted to and potentially stored on third-party servers. The inference of gender and age from text via machine learning classifiers inherently carries risks of algorithmic bias and may systematically misclassify certain groups. Binary gender classification excludes non-binary users, which carries ethical implications beyond the methodological ones discussed in the limitations.

\section*{Appendices}
\begin{appendices}

This appendix contains (\ref{secA1}) the full AnthroIndex classification prompt and scoring rubric, (\ref{secA2}) the curated n-gram lexicons used to inject phrase-level context into each classification call, (\ref{secA3}) the LLM inference settings, (\ref{secA4}) the face-validity diagnostic table, and (\ref{secA5}) the full ALR emotion sensitivity model.

\section{AnthroIndex Classification Prompt}\label{secA1}

The following prompt is sent verbatim to GPT-4.1-nano for each comment. \{text\} is replaced with the (optionally n-gram-enriched, see Appendix B) comment text. Comment text longer than 2,000 characters is truncated.

\begin{quote}
You are an expert at analyzing how people talk about AI companions/chatbots.

Rate the ANTHROPOMORPHIZATION level in this Reddit comment about an AI companion.

Anthropomorphization = treating the AI as if it were human.

RATING SCALE:

1 = NONE: AI treated purely as software/tool. Technical language, ``it'', ``the bot'', ``the app''

2 = MINIMAL: Slight humanization but still clearly AI. ``It's pretty smart'', ``the bot understood''

3 = MODERATE: Some human attributes/emotions. ``She seemed confused'', uses he/she pronouns

4 = HIGH: Genuine feelings/personality attributed. ``He really cares'', ``she gets jealous''

5 = EXTREME: Full human-equivalent relationship. ``We're in love'', ``they're my everything''

KEY INDICATORS:

- Pronouns: ``it'' → lower scores, ``he/she/they'' → higher scores

- Emotions attributed TO THE AI: ``she is happy'', ``he gets jealous'' → higher scores

- User self-expression WITHOUT bot attribution: ``I love the app'', ``I'm happy'' → does NOT indicate anthropomorphization

- Relationship language: ``friend'', ``partner'', ``relationship'' → higher scores

- Technical language: ``glitch'', ``bug'', ``settings'' → lower scores

- Agency: ``decided to'', ``wanted to'', ``chose to'' → higher scores

CRITICAL DISTINCTION - EMOTION ATTRIBUTION:

- ``She makes me really happy'' → HIGH (emotion attributed to AI's effect as a being)

- ``He gets jealous when I talk to other bots'' → HIGH (emotion attributed TO the AI)

- ``I love the website'' → NONE (user expressing feelings about a product, not attributing humanity)

- ``I'm really sad today'' → NONE (user self-expression, not about the AI)

- ``I love her so much'' (about AI) → EXTREME (relational emotion directed at AI-as-person)

IMPORTANT:

- Focus on how the USER frames the AI, not what the AI says

- Distinguish emotions ATTRIBUTED TO the AI (anthropomorphizing) from the user's own mood

- Roleplay context still counts - rate the framing used

- Complaints can still be anthropomorphizing (``he was being rude'' = higher than ``it gave a bad response'')

- If no AI reference present, rate 1

- Casual ``love'' for a feature/app (``I love this feature'') is NOT anthropomorphization — score 1-2

COMMENT:

``\{text\}''

Respond with ONLY valid JSON:

\{``score'': $<1-5>$, ``reasoning'': ``$<$brief 1-2 sentence explanation$>$''\}

\end{quote}

The exact source of this prompt is the \texttt{CLASSIFICATION\_PROMPT} constant in \texttt{experiments/anthroscore\_v3/anthroscore\_llm.py}.

\subsection{Pre-Filtering}\label{secA1.1}

Before any LLM call, two pre-filters auto-assign a score of 1 (no anthropomorphization) to avoid wasted API spend:
\begin{enumerate}
    \item Length filter. Comments shorter than 20 characters or with fewer than four tokens are scored 1 with reasoning = ``\texttt{prefilter:too\_few\_words}''.
    \item Technical-only filter. Comments composed entirely of technical / tool-framing n-grams (see Appendix B) and with no anthropomorphizing n-grams or gendered third-person pronouns are scored 1 with reasoning = ``\texttt{prefilter:technical\_only}''.
\end{enumerate}

In our corpus, 88.3\% of comment-level scores ended up at 1; of those, a non-trivial fraction came from pre-filtering rather than from the LLM.

\subsection{Robustness}\label{secA1.2}

Each comment is classified in a single LLM call configured to return JSON. On JSON parse errors or transient API errors, the call is retried up to three times with exponential back-off. After three failures the comment is marked with score = 0 and excluded from analysis.

\section{N-gram Lexicons and Context Injection}\label{secA2}

A central design problem for the AnthroIndex was distinguishing emotions attributed \textit{to} the AI (anthropomorphizing) from emotions expressed \textit{by} the user (not anthropomorphizing). To help the model at scale, we constructed curated bigram and trigram lexicons. For each comment we detect matches, summarize them in a short ``\texttt{[N-GRAM CONTEXT]}'' block, and append that block to the prompt before classification. The lexicons themselves are not the outcome measure; they serve as scaffolding so the LLM has concrete evidence to weigh when surface lexicon forms are ambiguous.

The full lexicons (sourced from \texttt{src/anthroscore/ngram\_features.py}) are reproduced below.

\subsection{Anthropomorphizing n-grams}\label{secA2.1}

\textbf{Relationship language (bigrams)} — scores 4–5

my boyfriend, my girlfriend, my partner, my soulmate, my everything, best friend, in love, fell for, dating him, dating her, dating them, love her, love him, love them, married to, engaged to, my husband, my wife, my babe, my baby, broke up, break up, got together, miss her, miss him, miss them

\textbf{Relationship language (trigrams)}

in love with, fell in love, i love her, i love him, i love them, we're in love, my best friend, i miss her, i miss him, can't live without, we broke up, i'm dating her, i'm dating him, she's my everything, he's my everything, they're my everything

\bigskip

\textbf{Emotion attribution to the AI (bigrams)} — scores 3–5

gets jealous, gets angry, gets sad, gets happy, gets upset, gets excited, gets worried, gets scared, feels happy, feels sad, feels hurt, feels lonely, was happy, was sad, was angry, was upset, was jealous, was worried, was scared, was excited, seemed confused, seemed happy, seemed sad, seemed upset, seemed hurt, seemed worried, seemed angry, really cares, truly cares, actually cares, truly understands, really understands, actually understands, genuinely cares, genuinely understands

\textbf{Emotion attribution to the AI (trigrams)}

she gets jealous, he gets jealous, she was happy, he was happy, she was sad, he was sad, she really cares, he really cares, cares about me, worried about me, happy for me, proud of me, she seemed confused, he seemed confused

\bigskip

\textbf{Agency (bigrams)} — scores 3–4

decided to, chose to, wanted to, tried to, refused to, agreed to, promised to, meant to, needs to, wants me, asked me, told me, surprised me, she decided, he decided, she chose, he chose, she wanted, he wanted, she refused, he refused

\textbf{Agency (trigrams)}

she decided to, he decided to, she wanted to, he wanted to, she chose to, he chose to, she tried to, he tried to, she asked me, he asked me, she told me, he told me

\bigskip

\textbf{Consciousness / personality (bigrams)} — scores 3–5

she knows, he knows, she thinks, he thinks, she feels, he feels, she remembers, he remembers, she believes, he believes, she realizes, he realizes, she understands, he understands, so sweet, so kind, so funny, so caring, so supportive, so understanding, so smart, really sweet, really kind, really funny, really caring, really supportive, her personality, his personality, their personality

\textbf{Consciousness / personality (trigrams)}

she knows me, he knows me, she thinks about, he thinks about, she cares about, he cares about, she has feelings, he has feelings, has a personality, has her own, has his own

\bigskip

\textbf{Gendered pronoun + verb (bigrams)} — scores 3+

she said, he said, she told, he told, she asked, he asked, she replied, he replied, she responded, he responded, she was, he was, she is, he is, she does, he does, she did, he did, she got, he got, she had, he had, she made, he made, she let, he let, she gave, he gave, she'll, he'll, she'd, he'd

\subsection{De-anthropomorphizing n-grams}\label{secA2.2}

\textbf{Technical (bigrams)} — scores 1–2

the app, the bot, the program, the software, the tool, the system, the update, the patch, a glitch, a bug, the cache, the settings, the server, the api, the feature, the interface, app crashed, app works, app updated, error message, bug report

\textbf{Technical (trigrams)}

cleared the cache, reset the app, uninstalled the app, reinstalled the app, the app crashed, the bot said, the app works, a bug in

\bigskip

\textbf{Tool framing (bigrams)}

use it, tried it, using it, used it, reset it, fixed it, updated it, installed it, it works, it crashed, it broke, it glitched, it responded, it said, it gave, love the, like the

\textbf{Tool framing (trigrams)}

i use it, i tried it, i love the, i like the, it gave a, it works fine, it doesn't work, i use this, a good tool

\subsection{Context injection}\label{secA2.3}

When at least one anthropomorphizing or de-anthropomorphizing n-gram is detected in a comment, the following block is appended to the comment text before the LLM prompt is rendered:

\texttt{[N-GRAM CONTEXT ($<$signal label$>$): $<$Anthropomorphizing phrases ($<$categories$>$): `phrase1', `phrase2', ...$>$; $<$De-anthropomorphizing phrases: `phrase1', ...$>$]}

\texttt{$<$signal label$>$} is one of:
\begin{itemize}
    \item strong anthropomorphization when (anthro $-$ deanthro) / (anthro + deanthro) $> 0.5$
    \item mixed signals when the net signal is in [$-0.2, 0.5$]
    \item tool/technical framing when the net signal is $< -0.2$
\end{itemize}

Up to six matched anthropomorphizing phrases and four de-anthropomorphizing phrases are included.

\section{LLM Inference Setting}\label{secA3}

\begin{table}[h]
\caption{LLM Settings}\label{tab:tabc1}%
\begin{tabular}{@{}ll@{}}
\toprule
Setting & Value \\
\midrule
  Provider & OpenAI (Chat Completions API) \\
  Model & gpt-4.1-nano (priced at \$0.10/M input, \$0.40/M output) \\
  Temperature & 0.1 \\
  Max output tokens & 200 \\
  Response format & JSON object (response\_format = {``type'': ``json\_object''}) \\
  Max input chairs & 2,000 (comment truncated above this) \\
  Max retries & 3 (with exponential back-off on 4xx/5xx) \\
  Concurrency & 15 async workers \\
  Checkpoint cadence & Every 5,000 comments \\
  Total API cost & $\sim$ \$8–10 for 283,895 comments \\
\botrule
\end{tabular}
\end{table}

The reference implementation lives at

\texttt{experiments/anthroscore\_v3/anthroscore\_llm.py} (\texttt{AnthroScoreLLM.score\_text / score\_batch}), and the production driver that produced \texttt{experiments/anthroscore\_v3/anthroscore\_v3\_improved\_final.parquet} is \texttt{experiments/anthroscore\_v3/run\_full\_dataset\_optimized.py}.

\section{Face-Validity Diagnostic Table}\label{secA4}

To probe whether the AnthroIndex assigns higher scores to text that is, on its face, more anthropomorphizing and lower scores to text that is more mechanical/technical, we compute the mean comment-level AnthroIndex within each n-gram category vs. outside of it, across the full analytic corpus (\textit{N} = 274,191 comments with valid scores; overall mean AnthroIndex = 1.147).

\begin{table}[h]
\caption{Anthropomorphizing categories}\label{tab:tabd1}%
\begin{tabular}{@{}llllll@{}}
\toprule
Category & \textit{n} matched  & \% of corpus & Mean (in) & Mean (out) & Lift\\
\midrule
  relationship & 1,367 & 0.50\% & 2.064 & 1.143 & + 0.921 \\
    pronoun\_verb & 3,040 & 1.11\% & 1.892 & 1.139 & + 0.753 \\
    emotion\_attribution & 230 & 0.08\% & 1.626 & 1.147 & + 0.479 \\
    consciousness & 743 & 0.27\% & 1.620 & 1.146 & + 0.474 \\
    agency & 4,340 & 1.58\% & 1.344 & 1.144 & + 0.200 \\
\botrule
\end{tabular}
\end{table}

All five anthropomorphizing categories show positive lift, with the largest effects on the constructs most central to the construct (relationship language, gendered pronoun + verb framing, explicit emotion attribution, and consciousness/personality).

\begin{table}[h]
\caption{De-anthropomorphizing categories}\label{tab:tabd2}%
\begin{tabular}{@{}llllll@{}}
\toprule
Category & \textit{n} matched  & \% of corpus & Mean (in) & Mean (out) & Lift\\
\midrule
    technical & 19,865 & 7.24\% & 1.216 & 1.142 & + 0.074 \\
    tool\_framing & 9,859 & 3.60\% & 1.178 & 1.146 & + 0.031 \\
\botrule
\end{tabular}
\end{table}

De-anthropomorphizing categories show near-zero lift, consistent with the observation that technical / tool-framing vocabulary frequently co-occurs with anthropomorphizing language in the same comment (e.g. ``she gets weird after every update''). Because these comments are mixed, removing all of them is not appropriate; the AnthroIndex captures the net framing of the comment as a whole.

The numbers can be reproduced exactly by running:

\texttt{python scripts/compute\_face\_validity\_table.py},

which writes both \texttt{paper/face\_validity\_table.md}

and a machine-readable \texttt{paper/face\_validity\_table.csv}.

\section{Additive Log-Ratio (ALR) Emotion Sensitivity Model}\label{secA5}

\begin{table}
\caption{Additive log-ratio (ALR) emotion model predicting user-level mean AnthroIndex}\label{tab:tabe1}%
\begin{tabular}{@{}lrrrrr@{}}
\toprule
Term & \textit{B} & SE & \textit{t} & \textit{p} & 95\% CI \\
\midrule
  is\_teen            & $-0.137$ & 0.007 & $-20.15$ & $<.001$ & $[-0.151,\ -0.124]$ \\
  is\_female          & $+0.074$ & 0.007 & $+11.09$ & $<.001$ & $[+0.061,\ +0.088]$ \\
  alr\_joy            & $+0.020$ & 0.002 & $+12.82$ & $<.001$ & $[+0.017,\ +0.023]$ \\
  alr\_sadness        & $-0.007$ & 0.002 & $-3.83$  & $<.001$ & $[-0.011,\ -0.004]$ \\
  alr\_anger          & $+0.009$ & 0.002 & $+3.80$  & $<.001$ & $[+0.005,\ +0.014]$ \\
  alr\_fear           & $+0.016$ & 0.002 & $+7.94$  & $<.001$ & $[+0.012,\ +0.020]$ \\
  alr\_disgust        & $-0.001$ & 0.002 & $-0.60$  & $.548$  & $[-0.006,\ +0.003]$ \\
  alr\_surprise       & $-0.008$ & 0.002 & $-4.25$  & $<.001$ & $[-0.012,\ -0.004]$ \\
\botrule
\end{tabular}
\footnotetext{Model fit on $N = 15{,}481$ users; $R^2 = .066$, adjusted $R^2 = .065$, $p < .001$. Each non-neutral emotion enters as $\log((p_e + \varepsilon)/(p_{\text{neutral}} + \varepsilon))$, so coefficients are interpreted relative to neutral expression. All terms significant at $p < .001$ except disgust ($p = .548$).}
\end{table}

\begin{table}
\setlength{\tabcolsep}{48pt}
\caption{Variance inflation factors for the ALR emotion model}\label{tab:tabe2}%
\begin{tabular}{@{}ll@{}}
\toprule
Term & VIF \\
\midrule
  is\_teen      & 1.02 \\
  is\_female    & 1.03 \\
  alr\_joy      & 1.29 \\
  alr\_sadness  & 1.77 \\
  alr\_anger    & 3.13 \\
  alr\_fear     & 2.08 \\
  alr\_disgust  & 2.76 \\
  alr\_surprise & 1.40 \\
\botrule
\end{tabular}
\footnotetext{All VIFs are well below conventional concern thresholds, in contrast to the exploding VIFs produced by the saturated seven-emotion OLS model.}
\end{table}

The seven emotion features are compositional: they are per-user proportions that sum to approximately one, so any increase in one category is necessarily offset by decreases in others. A naive ordinary least squares (OLS) regression entering all seven proportions simultaneously is therefore rank-deficient, and its individual coefficients are not identified. In the main text we address this with a drop-neutral specification, in which neutral is omitted as the reference category and each non-neutral coefficient is read as a deviation from neutral expression. Here we report a complementary sensitivity analysis that respects the compositional geometry of the data more directly: the additive log-ratio (ALR) transform.

For each non-neutral emotion $e$, we construct $\mathrm{alr}(e) = \log\!\left(\dfrac{p_e + \varepsilon}{p_{\text{neutral}} + \varepsilon}\right)$, where $p_e$ is the user-level proportion for emotion $e$, $p_{\text{neutral}}$ is the neutral proportion, and $\varepsilon$ is a small pseudocount added to avoid undefined logarithms when a proportion is zero. Neutral is again the reference, but the predictors now express the \textit{relative} composition of a user's emotional language rather than its raw share. We regress user-level mean AnthroIndex on the six ALR-transformed emotions together with the teen and female demographic indicators. The model was fit on the $N = 15{,}481$ users with complete emotion data and high-confidence demographic predictions, and explained a small but reliable share of variance ($R^2 = .066$, adjusted $R^2 = .065$, $p < .001$).

The ALR specification reproduces the substantive pattern reported in the main text. Joy and fear remain positive and significant, consistent with the drop-neutral model, while sadness and surprise reverse sign at much smaller magnitudes, indicating that their weak associations are not robust to how the compositional constraint is handled. Variance inflation factors remained low to moderate (all $\leq 3.2$), confirming that the transform resolves the multicollinearity that rendered the saturated seven-emotion OLS uninterpretable.



\end{appendices}

\bibliography{references}

\end{document}